\documentstyle[prl,floats,aps,twocolumn,epsf,graphicx]{revtex}
\begin{document}
\twocolumn[\hsize\textwidth\columnwidth\hsize\csname
@twocolumnfalse\endcsname

%\begin{document}
\title{Evolutionary Minority Game: The Roles of Response Time and Mutation Threshold}
\author{Shahar Hod and Ehud Nakar}
\address{The Racah Institute of Physics, The Hebrew University, Jerusalem 91904, Israel}
\date{\today}
\maketitle

\begin{abstract}

\ \ \ In the evolutionary minority game, agents are allowed to evolve 
their strategies (``mutate'')  based on past experience. We explore the dependence of the 
system's global behavior on the response time and the mutation threshold of the agents. 
We find that the precise values of these parameters determine if the strategy distribution 
of the population has a U-shape, inverse-U-shape, or a W-shape. 
It is shown that in a free society (market), highly adaptive agents (with short response times) 
perform best. In addition, ``patient'' agents (with high mutation threshold) 
outperform ``nervous'' ones.
\end{abstract}
\bigskip

]

A problem of wide interest in biological and socio-economic systems
is that of an evolving population in which individual agents adapt their
behavior according to past experience. The Minority Game (MG) is one of the 
most studied models of such complex systems (see e.g., 
\cite{HubLuk,ChaZha,DhRo,SaMaRi,John1,Cev,BurCev,LoHuJo,HuLoJo,HaJeJoHu,LoLiHuJo,LiVaSa,BurCevPer,BuCePe,HodNak,NakHod,Hod,MetChr,SouHal,Min} and references therein). 
In this model, a population of $N$ agents with limited information and capabilities repeatedly 
compete for a limited global resource, or to be in the minority. 
In financial markets for instance, more sellers than buyers implies lower prices, 
and it is therefore better for a trader to be in a minority group of buyers. 
Predators foraging for food will do better if they hunt in areas with fewer competitors. 
Rush-hour drivers, facing the choice between two alternative routes, wish to choose 
the route containing the minority of traffic \cite{HubLuk}. 

At each round of the game, every individual has to 
choose whether to be in room `0' (e.g., choosing to sell an asset or taking route A) 
or in room `1' (e.g., choosing to buy an asset or taking route B). 
At the end of each turn, agents belonging to the smaller group (the minority) are 
the winners, each of them gains one point (the ``prize''), 
whereas the others lose a point (the ``fine''). 
The agents have a common `memory' look-up table, containing the outcomes of 
recent occurrences. 
Faced with a given bit string of recent occurrences, each agent chooses 
the outcome in the memory (the so-called  ``predicted trend'') 
with probability $p$, known as the agent's ``gene'' value (and the opposite alternative with 
probability $1-p$). 

The evolutionary formulation of the model (EMG) 
\cite{John1,HodNak} allows agents to adapt their strategy according 
to their past experience: if an agent score falls below some value $D$ 
(the mutation threshold), he mutates -- 
its gene value is modified. In this sense, 
each agent tries to learn from his past mistakes, 
and to adjust his strategy in order to survive. 

In previous studies of the EMG, the criterion according to 
which each agent decided whether or not to change his strategy was based on his performance 
in {\it all} previous rounds of the game, giving {\it equal} weights to each of these rounds. 
Such a crude criterion lacks the capability of quantifying the 'local' performance of an agent 
(his net success in the last few rounds of the game). 
It may therefore lead to situations in which agents are taking the wrong 
decisions (based on the state of the 
system in the far past) without noticing that the system has already 
evolved into a completely different global state. Thus, of great interest for the study of 
realistic systems of competing (and evolving) agents are situations in which agents are capable of adapting their 
strategy according to their present ({\it local}) performance (rather than using a crude criterion for 
mutation, a one which gives equal weights to all previous rounds of the game).

The aim of the present work is to explore the dynamics of evolving populations with 
various levels of adaptation (various response times, see a precise definition below) 
and with different values of the mutation threshold. Of main importance 
is the identification of the strategies that perform best in a particular situation. 

In the present formulation of the model, each agent holds a measure of his past performance through a 
moving average $S(t;T)$, whose value reflects the payoffs from recent $T$ 
rounds of the game. The moving average is updated with each turn of the game \cite{KayJohnBenj}:

\begin{equation}\label{Eq1}
S(t;T)={{T-1} \over T} S(t-1;T) + {1 \over T} \Delta(t) \  ,
\end{equation}
where $\Delta(t)=\pm 1$ is the agent's payoff at time step $t$. Thus, the information about 
previous outcomes has a 'half life' of $\sim T$ turns [the contribution of a given turn to $S(t;T)$ falls 
exponentially with successive rounds]. If the moving average of an agent falls below the mutation 
threshold, his strategy (i.e., its gene value) is modified. After mutation, the agent enters a 
``trial period'' of $T$ rounds before considering mutating again. The mutation threshold $D$ characterizes the 
``patience'' of an agent. The smaller is the value of $D$ the more tolerance 
(willing to suffer some local losses without modifying his strategy) is 
the agent. The value of the parameter $T$ is a measure of the agent's 
level of adaptiveness, his response time to temporal changes in the state of the system. 
The smaller is the value of $T$, the faster is the 
agent's response to any deterioration in his performance.

\begin{figure}[tbh]
\centerline{\epsfxsize=9cm \epsfbox{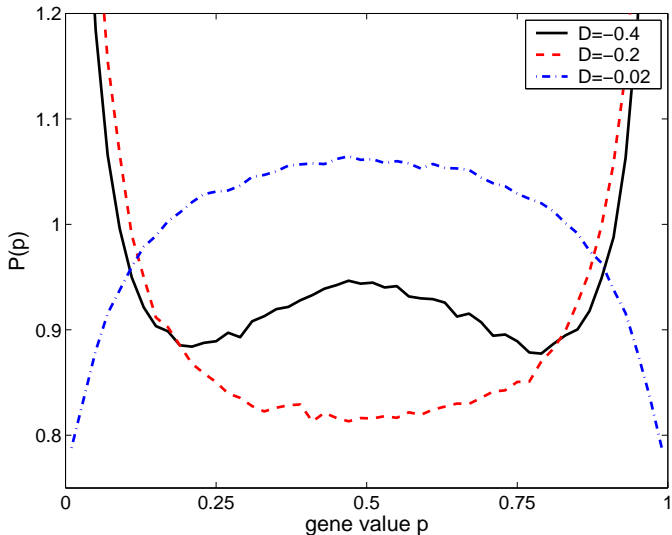}} 
\caption{The strategy distribution $P(p)$ for different values of the mutation threshold. 
The results are for $N=10001$ agents, and a fixed response time of $T=25$. Each point represents 
an average value over 10 runs and 20000 time steps per run.}
\label{Fig1}
\end{figure}

Figure \ref{Fig1} displays the long-time averaged gene distribution $P(p)$ of the agents for 
a fixed response time. 
We find three qualitatively different populations, depending on the precise value of the 
mutation threshold $D$. For $D < D^{(1)}_c$ (this corresponds to a population of ``patient'' agents, ones who are willing to suffer some temporary 
losses without changing their strategies) 
the population tends to form a W-shaped distribution (The precise value of $D^{(1)}_c$ depends on the value of the response time $T$). Remarkably, we find that 
this W-shaped strategy distribution is dynamically meta-stable. One observes that from time to time the 
system undergoes a short and abrupt change into an inverse-U 
shaped distribution (which quickly returns to a W-shaped distribution). 
On the other hand, for $D > D^{(2)}_c$ 
(``nervous'' agents who hurry to change their strategies due to even small local losses) 
the population tends to crowd around $p={1 \over 2}$, forming a (stable) inverse-U shaped gene distribution. 
This corresponds to ``confused'' and ``indecisive'' agents (agents that 
prefer a coin-tossing strategy). There is also an intermediate phase [for $D^{(1)}_c < D < D^{(2)}_c$], in which $P(p)$ has a U-shape 
with two symmetric peaks at $p=0$ and $p=1$ -- the population tends to self-segregate 
(this corresponds to always or never following what happened last time). To 
flourish in such a population, an agent should behave in an {\it extreme} way. 

\begin{figure}[tbh]
\centerline{\epsfxsize=9cm \epsfbox{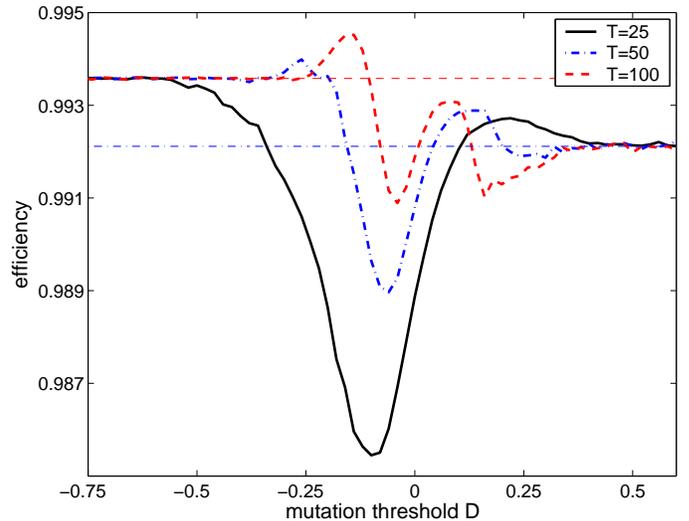}} 
\caption{The efficiency of the system as a function of the mutation threshold $D$. 
Horizontal lines represent the efficiency for uniform $P(p)$ distribution (dashed) and 
a coin-tossing situation (dashed-dotted). Initially, there is a uniform distribution of the strategies. 
The results are for $N=10001$ agents. Each point represents 
an average value over 10 runs and 20000 time steps per run.}
\label{Fig2}
\end{figure}

The (scaled) efficiency of the system is defined as the number of agents in the 
minority room, divided by the maximal possible size of the 
minority group, $(N-1)/2$. 
Figure \ref{Fig2} displays the system's efficiency as a function of the mutation threshold 
$D$ (and for various different values of the response time $T$). We also display the 
efficiency for agents guessing {\it randomly} between room `0' and room `1', and for a uniform distribution of agents. There is a range of mutation thresholds $D$ for 
which the efficiency of the system is {\it better} than the random case. Thus, the agents 
cooperate {\it indirectly} to achieve an optimum utilization of the system's resources. 
However, there is also a range of $D$ values for which 
the efficiency of the system is remarkably {\it lower} than that obtained for agents choosing via 
independent coin-tosses. Thus, considering the efficiency of the system as a whole, 
the agents would be better off not adapting their 
strategies because they are doing {\it worse} than just guessing at random.

\begin{figure}[tbh]
\centerline{\epsfxsize=9cm \epsfbox{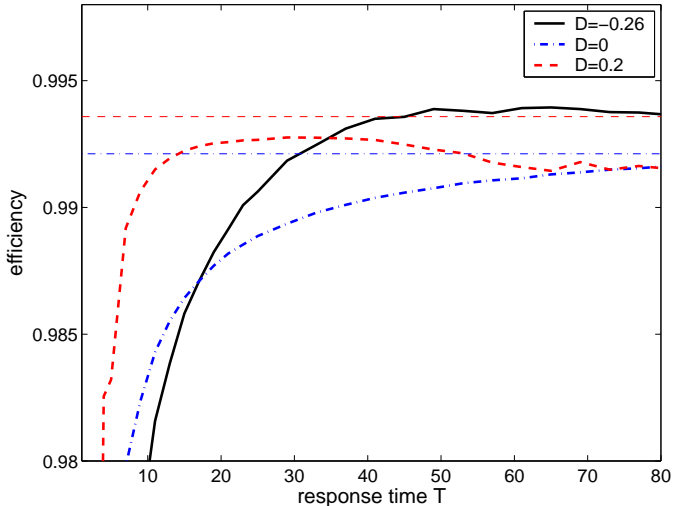}} 
\caption{The efficiency of the system as a function of the (common) response time $T$. 
Horizontal lines represent the efficiency for uniform $P(p)$ distribution (dashed) and 
a coin-tossing situation (dashed-dotted). Initially, there is a uniform distribution of the strategies. 
The results are for $N=10001$ agents. Each point represents 
an average value over 10 runs and 20000 time steps per run.}
\label{Fig3}
\end{figure}

Figure \ref{Fig3} displays the system's efficiency as a function of the response time 
$T$ (and for various different values of the mutation threshold $D$). Note that the 
system's global efficiency is a monotonic increasing function of the response time for intermediate values of the 
mutation threshold. However, for systems composed of nervous agents (large $D$ values), and for 
systems composed of patient members (very small $D$ values), 
the utilization of the system's resources is optimal for intermediate response times.

\begin{figure}[tbh]
\centerline{\epsfxsize=9cm \epsfbox{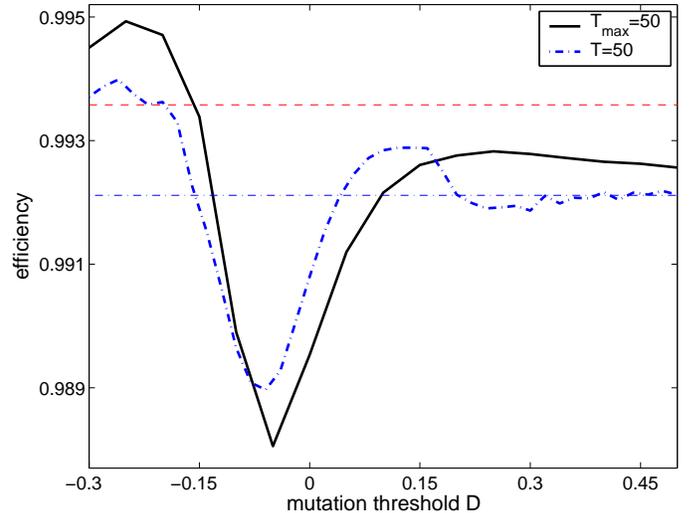}} 
\caption{The efficiency of the system as a function of the (common) mutation threshold $D$. 
Agents have a common mutation threshold, but different response times $1 \leq T_i \leq 50$. For 
comparison we also display the efficiency of a system composed of agents with a common 
response time of $T=50$ (dashed-doted curve). 
Horizontal lines represent the efficiency for uniform $P(p)$ distribution (dashed) and 
a coin-tossing situation (dashed-dotted). 
The results are for $N=10001$ agents. Each point represents 
an average value over 100 runs and 20000 time steps per run.}
\label{Fig4}
\end{figure}

Next, we relax the condition that all members have the same (common) response time. We consider a 
population of competing and evolving agents in which each individual is free
to adapt a personal response time in a range $1 \leq T_i \leq T_{max}$. 
Figure \ref{Fig4} displays the system's efficiency 
as a function of the mutation threshold (which is still common to all members of the 
population). For comparison, we also display the efficiency of an homogeneous population in which 
all agents have the same response time. 
One finds that allowing each agent to choose his own personal response time may 
{\it improve} the global efficiency of the system. Note however, that for 
intermediate values of the mutation threshold, this freedom (to choose a personal response time) may cause a decrease in the system's global efficiency.

\begin{figure}[tbh]
\centerline{\epsfxsize=9cm \epsfbox{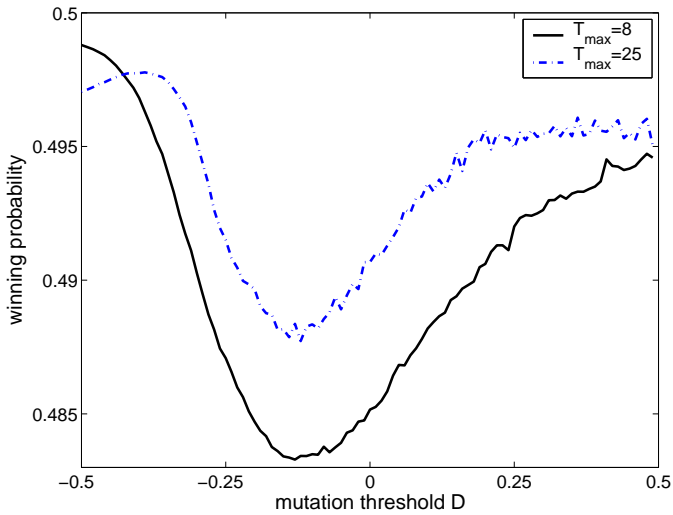}} 
\caption{The winning probability of an agent as a function of his mutation threshold. 
Each agent is free to adapt a personal mutation threshold and a personal response time. 
The results are for $N=10001$ agents. Each point represents 
an average value over 100 runs and 20000 time steps per run.}
\label{Fig5}
\end{figure}

Finally, we consider the case of a free society (market) in which each 
member is allowed to choose both his personal response time and his mutation threshold as well. 
Figure \ref{Fig5} displays the winning probability of an 
agent in such a population as a function of his personal mutation threshold. We find that agents with 
small (negative) values of the mutation threshold $D$ perform best. These are ``patient'' agents who are willing to 
suffer some temporary losses without modifying their strategy.

\begin{figure}[tbh]
\centerline{\epsfxsize=9cm \epsfbox{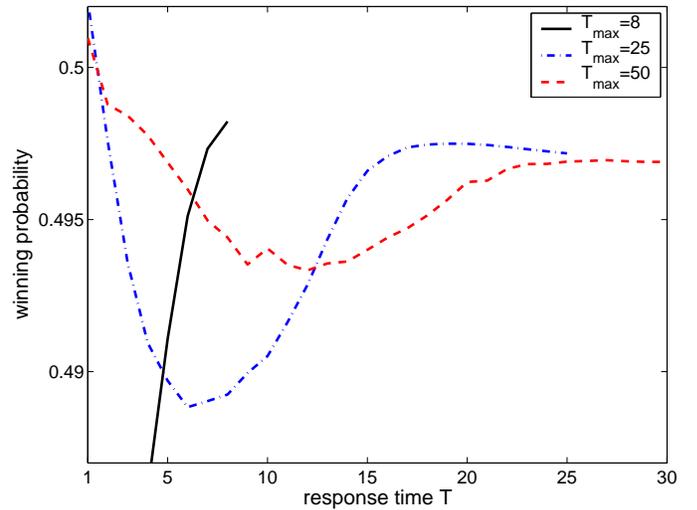}} 
\caption{The winning probability of an agent as a function of his 
response time. Each agent is free to adapt both a 
personal mutation threshold and a personal response time. 
The parameters are the same as in Fig. \ref{Fig5}.}
\label{Fig6}
\end{figure}

In Fig. \ref{Fig6} we display the winning probability of an agent 
as a function of his personal response time. One finds that in such free populations agents with short response times perform best. 
In fact, their winning probability {\it exceeds} $50\%$. It turns out that these agents asses their performance 
very often, which allows them to respond quickly and efficiently to any change in the global state of the system. 
On the other hand, the winning probability has a minimum at intermediate values of the response time. 
[Note however, that in a population composed of agents with only short 
response times ($T_{max}=8$ in Fig. \ref{Fig6}), 
it is best to have the largest response time available].

In summary, we have explored the dynamics of complex adaptive systems with various 
different values of response times and mutation thresholds. 
The main results and their implications are as follows:

(i) A population of ``patient'' agents [$D < D^{(1)}_c$] tends to form a W-shaped 
distribution of strategies. The W-shaped gene-distribution is intriguing in the sense that it does 
not appear in adaptive systems in which agents asses their performance according to {\it all} 
previous rounds of the evolution \cite{John1,HodNak}. This is a new feature of the 
present model.

On the other 
hand, a population of ``nervous'' agents [$D > D^{(2)}_c$] tends 
to cluster around $p={1 \over 2}$ (a coin-tossing strategy). 
Stated in a more pictorial way, confusion and indecisiveness take over in nervous systems.

(ii)  An evolving population achieves an 
{\it optimum} utilization of its global resources for small negative values of the 
mutation threshold $D$ (see Fig. \ref{Fig2}). This corresponds to a 
population of patient members. 
For large $D$ values agents tend to be indecisive (preferring a coin-tossing strategy), 
a behavior which destroys any attempt to establish (indirect) cooperation. 
It seems that ``nervousness'' prevents the agents from achieving a reasonable utilization of their resources.

(iii) In a {\it free} society of competing agents (in which each member has the freedom to 
adapt his own response time and mutation threshold) 
patient agents perform best (see Fig. \ref{Fig5}). 

(iv) The best performance is achieved by 
agents who have very short response times (see Fig. \ref{Fig6}). 
These agents have a high level of 
adaptiveness, making it possible for them to response quickly and efficiently 
to local changes in the state of the system. 
The success rate of such agents actually exceeds $50\%$. 
(Agents who have very long response times also perform reasonably well, 
whereas the winning probability drops to a minimum at intermediate values of the response time). 

\bigskip
\noindent
{\bf ACKNOWLEDGMENTS}
\bigskip

The research of SH was supported by G.I.F. Foundation. We would like to 
thank Uri Keshet for his assistance.

\end{document}